# Social Networks as a Collective Intelligence: An Examination of the Python Ecosystem

By Thomas Pike[a], Robert Colter[a], Mark Bailey[a], Jackie Kazil[b] and John Speed Meyers[c]

[a]NIU – Washington DC, [b]Rebellion Defense – Washington, DC, [c]ChainGuard – Arlington, VA

**Abstract:**

The Python ecosystem represents a global, data rich, technology-enabled network. By analyzing Python's dependency network, its top 14 most imported libraries and cPython (or core Python) libraries, this research finds clear evidence the Python network can be considered a problem solving network. Analysis of the contributor network of the top 14 libraries and cPython reveals emergent specialization, where experts of specific libraries are isolated and focused while other experts link these critical libraries together, optimizing both local and global information exchange efficiency. As these networks are expanded, the local efficiency drops while the density increases, representing a possible transition point between exploitation (optimizing working solutions) and exploration (finding new solutions). These results provide insight into the optimal functioning of technology-enabled social networks and may have larger implications for the effective functioning of modern organizations.

      The structure and dynamics of social networks - from families to organizations to governments - can be understood as self-optimizing problem solving networks approaching a configuration that allows them to share information and generate solutions to enable their survival [1]–[3]. Capturing data to understand how these networks function presents challenges not only because of the difficulty of identifying nodes and edges but also because of issues in understanding how information flows across those edges and how the nodes process these flows of information. Identifying and analyzing data-rich, emergent systems can provide a critical resource to find insights into the dynamics of problem solving networks. By conducting an examination of the Python dependency network, its 14 most imported libraries, and cPython (the lower level language that creates Python) this research found strong evidence that social coding repositories can be understood as emergent collaborative and adaptive problem solving networks, or collective intelligences that are crucially data rich. In addition, these crowd sourced collective intelligences have demonstrated a clear competitive advantage in producing the software that fuels our modern world, while challenging traditional economic thinking [4], [5]. An initial network assessment of the Python dependency and contributor network revealed that Python has an emergent specialization where a small number of libraries are maintained by a small number of experts who specialize in that library. These crucial libraries are then connected together by enablers that allow these critical parts of the Python ecosystem to optimize both their local and global information efficiency.

These results demonstrate how the top libraries of the Python ecosystem balance the explore-exploit tradeoff to exploit existing solutions while also exploring to find new solutions.

## Background

The Python ecosystem is one part of a larger phenomenon of Open Source Software (OSS). OSS makes up a significant portion of the infrastructure that fuels our technical lives from the operating system to coding languages. OSS is defined as software code that is available for review, extensions or enhancements by anyone with the technical competency and desire to contribute.[6] A significant number (estimates vary from above 50% to 80%) of IT applications and infrastructure use at least some open source code.[7] The Python programming language is itself open source and is supported by an array of individuals, businesses, and foundations. OSS is made possible by the Internet and its ability to allow decentralized individuals to collaborate on projects, as well as the practical need for code to be extensible and reusable. This collaboration is enhanced by software (most notably git) that allows collaborators to rapidly review and merge code, and (critically for research) track and store records of changes. These dynamics have created a new paradigm of knowledge sharing and collaboration, while also capturing large amounts of data on collaboration dynamics.

The OSS ecosystem can be understood as a taxonomic tree of sub-ecosystems related to specific efforts (e.g. R, Linux, Python, etc.). Although a complete description of these efforts goes beyond the scope of this study, a brief discussion of some of the major parts is helpful in understanding OSS. First, OSS groupings can be associated with critical underlying infrastructure projects. This includes the Linux operating system, which has a wide range of open source and proprietary versions and a robust foundation that helps its development.[8] Or, the Apache Software foundation, which manages over 350 projects, to include the world's most used web software service.[9] Second, there are ecosystems associated with coding languages. These include the focus of this study: Python and the associated package managers like the Python Package Index (PyPI). Most major languages have a robust ecosystem of package managers such as Maven with Java, Node Package Manager (npm) with JavaScript, or RubyGems with Ruby. These ecosystems are crucial to each language as they provide users with documented code they can integrate into their specific coding projects, thereby reducing the associated costs and, if the package is well developed, reducing error.

Constantly rewriting the same is time consuming and introduces unnecessary errors. Reducing these costs of development crowdsourcing solutions allows for efficiencies and more effective exploration to develop new knowledge[1], [10]. Critically, the network dynamics of how these OSS ecosystems work and how new innovations enhance them, such as the Internet and version control, can provide significant new insights into how we can function as organizations and a society. Recognizing that OSS is a new paradigm in knowledge sharing that emerged endogenously, this study starts the journey to understand the emergent structure of OSS networks for their own sake. Understanding the dynamics of such structures has implications for improving organizational dynamics, finding insights into complex adaptive systems, and understanding the information technology revolution as it relates to improved knowledge sharing and collaboration.

## Related Work

The constantly collected and stored data associated with OSS development has produced rich datasets. Not surprisingly, there is a lot of research that examines OSS dynamics from different perspectives. There is even an annual international conference, *Mining Software Repositories,* that has occurred for nearly two decades and spans numerous different research questions related to OSS.[11] Inside the expansive research is a much smaller subset related to the network topology of different coding ecosystems. These studies can be generally categorized into two groups: studies that look at the network topology of dependency graphs (how often a package is used by another program), and studies that look at the contributor network (the developers who build these packages). Of the two, the former subset is larger. Particular studies of note include [6], which compares the topology network of CRAN (R package manager), PyPI (Python package manager), and npm (JavaScript package manager). Of interest from this work is the finding that the dynamics of the programming language help shape the ecosystem dynamics. Among R, Python, and JavaScript, Python has the largest standard library of functions, with a dependency graph that is more isolated, while npm lacks a standard library so it has the largest package manager index with many small packages that provides this basic functionality and is more connected.[12]  A similar study of note is [7], which compares JavaScript, Ruby,  and Rust and examines their evolution over time. This study confirmed the npm analysis of [6] and finds the result, also seen in this study, that a small subset of packages dominate the dependency graph. With the additional assessment that over time, the possibility of ecosystem collapse diminishes as more redundancy emerges.[13]

In the second category of contributor networks research there was one project of note. Depsy.org, a multi-year project, (which is no longer maintained) sought to get researchers who contribute to critical coding projects credit in a manner analogous to academic citations.[14] This study acknowledges the importance of code as stored knowledge in a more dynamic way than publications, while confronting academia's historic focus on publication. Critical to this study is the understanding that code is stored knowledge and the OSS dynamics allowed for a detailed record of how this collective intelligence functions. Depsy.org, however, did not publish comprehensive results but instead provided data that was used by other researchers to explore the contributor network.

Research using the Depsy.org dataset combines analysis of the dependency networks and the contributor networks. From this combined category, two research studies were relevant. First, [9] which used network analysis of the dependency and contributors networks to assess indirect and direct knowledge spillover effects. The results found that network structure was positively associated with project success. Specifically, project closeness centrality was associated with project success while contributor closeness centrality was not [5]. This is important as it identifies a network metric of collective problem solving and identifies a metric correlated with increased probability of success. The other relevant research consisted of [10] and [11], with [11] being a continuation of [10] examining the factors that impact OSS development. Both [10] and [11] use the Depsy.org data for their research and find that the more derivative a package is (meaning the more dependencies it has), the less likely it is to have a high impact. In addition, out-degree, closeness and pagerank centrality are correlated with project impact as found in [11]. Although this study does not contradict their findings, there are some concerns regarding the methodology. In [11], the authors remove outliers from the data. This study argues that the outliers are instead critical components that must be understood to appreciate how the collective intelligence of the ecosystem functions efficiently and effectively. Numerous studies have examined the network dynamics of OSS. Several of these studies tangentially identify critical insights into understanding how the network topology of OSS ecosystems can both aid and produce vulnerabilities to the functioning of the ecosystem as a collective intelligence.

This study adds to the literature of social networks, mining software repositories, and open source software literature by specifically focusing on understanding the emergent network topology of the Python ecosystem to assess how the ecosystem functions as a collective intelligence. This study will specifically describe the emergent structure and examine the main contributors of this ecosystem. The results provide insights into how organizations should function as a network to maximize their explore-exploit paradigm of efficiency and innovation. To accomplish this, this study proceeds in three parts. First, we present an overview of the methods and data used to conduct the study. Second, we discuss the results. Third, we propose next steps to further this research, given that this analysis only scratches the surface of a rich research area.

## Data and Methods

This study used Github data harvested by the GHTorrent project [15] to analyze the emergent network structure of the Python ecosystem. Github, with its large and diverse user base and the collection of critical Python codebases primarily hosted on its servers (not to mention hosting core Python itself), is an ideal dataset to understand both the Python dependency network and part of the contributor network. The GHTorrent Project is a project dedicated to the capture and storage of user transactions from GitHub's REST API, through the use of donated access tokens, provided the GitHub data of the Python ecosystem [15]. The data used was the GHTorrent June 2019 MySQL database, which included GitHub user transactions from October 2012 through June 2019. All data was retrieved and pulled from this custom server through the use of a combination of MySQL Workbench and custom Python code.[1]

The GHTorrent Project uses a complex schema to store and retrieve information on all types of user interactions, both with stored repositories of code and between GitHub users [15]. However, for this study, only a small portion of that schema was considered relevant. Specifically, four tables. (1) *users* (storing unique users and their properties), (2) *projects* (storing unique repositories and their properties, including data on what repo they are forked from, if any), (3) *commits* (storing who committed changes to a repo and when), (4) *pull_requests* (storing who requested changes be committed to a repo and whe, as well who responded to the request and how). The relationships between these four tables are shown in Figure 1.

---

[1] The code used in this study is available at
https://github.com/Open-Source-Software-Neighborhood-Watch/dynamics_python_ecosystem.git

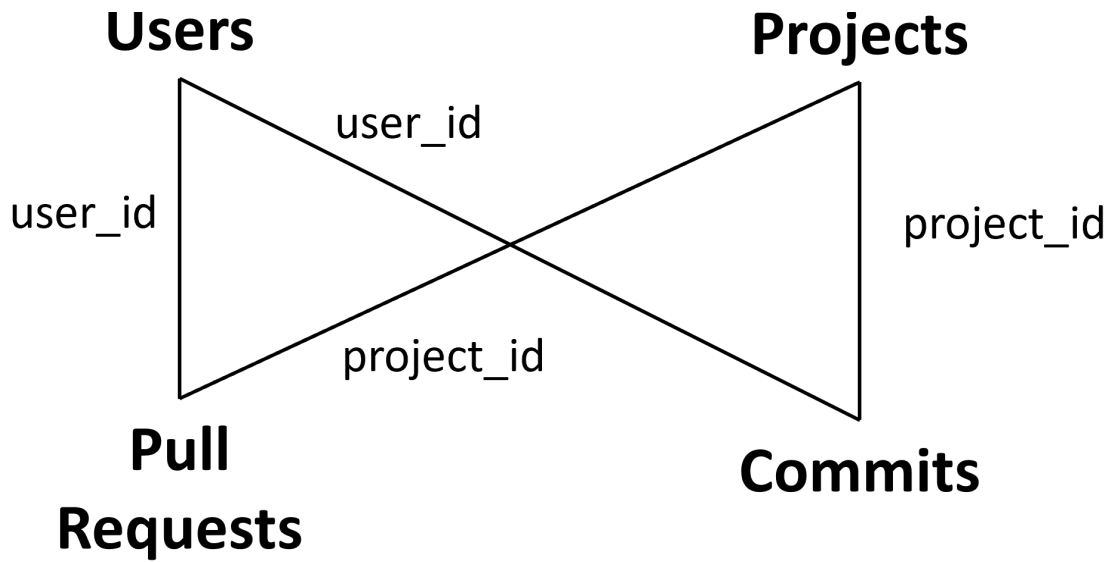

Figure 1: Database Schema

A customized query from the GHTorrent dataset provides the dependency and contributor data this study used to analyze the emergent network structure of the Python ecosystem. The quantitative approach was conducted in two phases. The first phase, which used statistical analysis to categorize the Python ecosystem, provided significant evidence that the ecosystem is in fact a complex adaptive system. This then provided the lens with which to apply and understand the network metrics. Finally, a grounded theory approach was used to assess how these social network dynamics aided the problem solving nature of the ecosystem as a complex system [16]. By categorizing the Python ecosystem and assessing its network metrics, this study generates an initial set of hypotheses about optimal social network structures for technology enabled collective problem solving.

## Descriptive Statistics

The statistical results of the dependency and contributor data show evidence of being complex adaptive systems based on the presence of heavy tailed distributions at multiple levels of both networks [17]–[20]. Both the dependency networks and the sample of contributor networks demonstrated heavy tailed dynamics. The Python dependency data comprises those libraries that were imported by at least one other repository. Applying this criteria results in a significant reduction of libraries from 165,027 validated libraries to 46,548 that import other libraries and 19,987 libraries which are imported. As the two lists are not mutually exclusive, the final graph resulted in a total of 59,311 nodes that make up the dependency graph. These 59,311 libraries then represent the critical dynamics this study endeavors to understand, how knowledge is developed and propagated across a decentralized, technology-enabled network. To assess the dynamics of the dependency network, histograms were created for the dependency data based on the number of times a library was imported. A graph was also created that linked any imported library to the importing code. As both the dependency data and is degree distribution corollary were heavy-tailed, the types of distributions were then assessed using the log-likelihood ratio test with the result that the dependency data is most similar to a truncated power law and the degree distribution moist similar to a power law[21], [22] (figure 2).

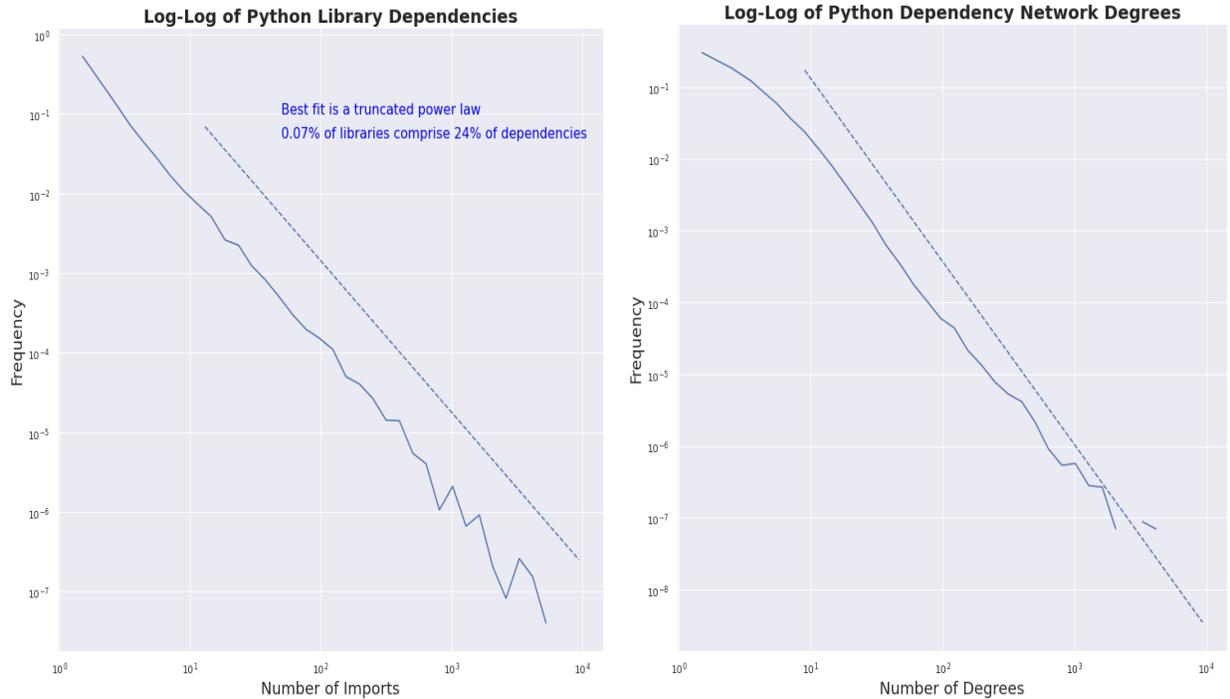

Figure 2: Log-likelihood Test Results of Dependency data(left) and degree distribution (right). Dependency data is defined as the number of times a Python library is imported by other code. Degree distribution is the number of degrees of each node.,

Due to the challenges of obtaining contributor data for all the libraries this study exploited the heavy tail nature of the ecosystem. By focusing on the tail of the distribution this research is able to focus on the contributor network of a small number of libraries and get insights into the contributor network dynamics that provide the majority of contributions to the functioning of the ecosystem. The contributor data is the number of contributions by a unique username to a unique library. Although a rough metric, a contribution is defined as the count of a pull request or commit. The tail of the distribution was identified by using the Paretian binning strategy proposed by Jiang [23]. Through this process the tail bin composes 14 Python Libraries (figure 3).

After identifying the tail of the distribution, the authors then retrieved the contribution data (commits and successful pulls) by user for each library. This data then forms the contributor network for the 14 most imported Python libraries. Each contributor network represents a collaborative project, where a team of individuals are working together to solve a problem. By adopting this approach we arguably assess the most successful libraries in the Python ecosystem. Quantitatively stated, by analyzing just 14 libraries we cover 24% of those libraries which are imported. Of these 14 libraries, there is great variance in the contribution network. Contributions ranged from 48 (setup tools) to 301,920 (odoo). Pytest-cov was nothing but commits, while setuptools was nothing but pull requests. In addition, some packages had thousands of contributors with Django having the most with 1948, while others only had a few, with setuptools having the least with two. This variance in itself is intriguing as it implies useful solutions can range from small libraries with a specific purpose and a small number of contributors to highly complex libraries with wide functionality and lots of contributors.

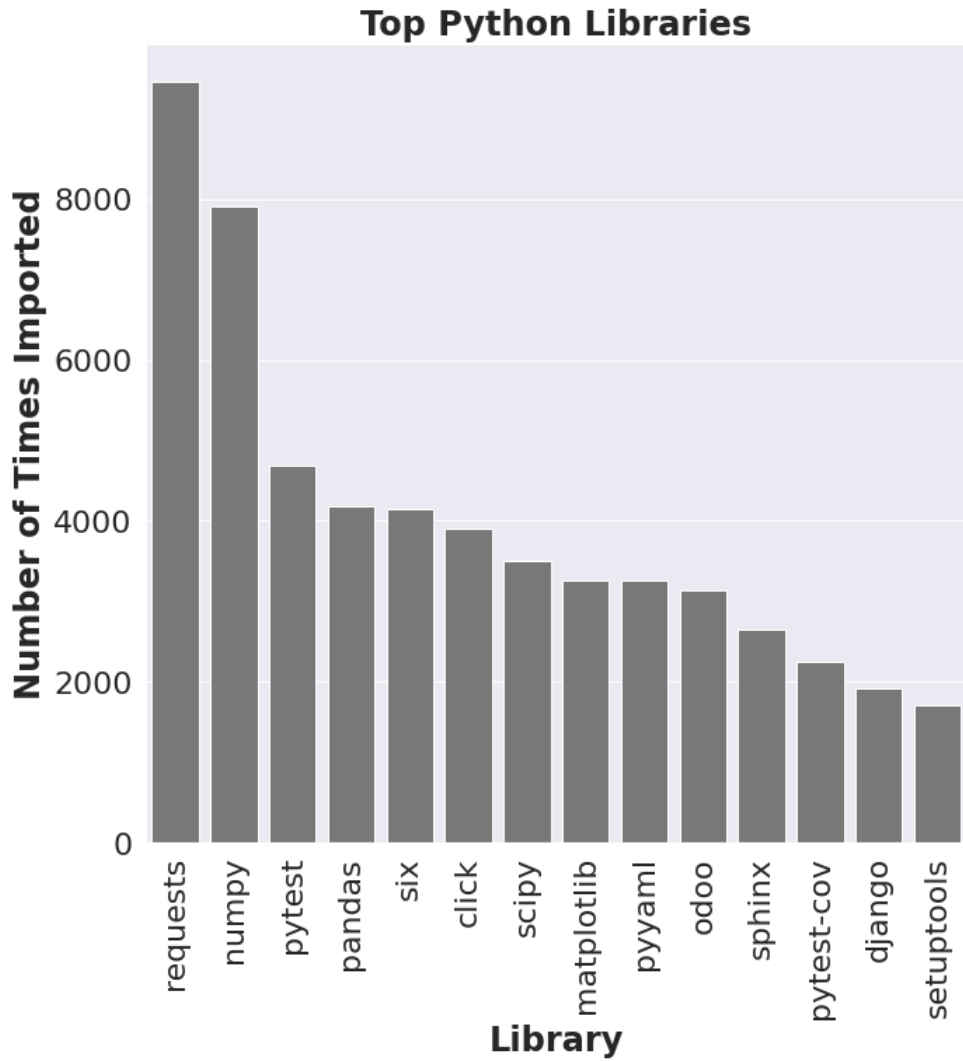

Figure 3: Histogram of the 14 most imported Python libraries. 14 was determined using the Paretian Binning Strategy [23] to identify the tail of the distribution.

As setuptools only consisted of two contributors, this library was then removed from the following calculations. Each contribution network was then examined the same way as the dependency network by assessing the distribution of contributions. The degree distribution was not assessed as each library is an ego network with all contributors tied to one library, their projections are examined later. After first visually assessing each library's contribution distribution character and determining heavy-tailed, the log likelihood ratio test was then used. Each contributor network then showed a distribution of either lognormal or truncated power law (table 1).

| Repository Name | Commits | Successful Pulls | Best Fit Distribution |
|---|---|---|---|
| cPython | 34159 | 566 | truncated_power_law |
| click | 592 | 16 | lognormal |
| django | 75720 | 658 | lognormal |
| matplotlib | 44068 | 286 | lognormal |
| numpy | 32502 | 126 | lognormal |
| odoo | 290904 | 11016 | truncated_power_law |
| pandas | 27534 | 156 | lognormal |
| pytest | 6932 | 110 | lognormal |
| pytest-cov | 598 | 0 | lognormal |
| pyyaml | 626 | 12 | truncated_power_law |
| requests | 5352 | 150 | lognormal |
| scipy | 24450 | 46 | truncated_power_law |
| six | 140 | 12 | lognormal |
| sphinx | 20700 | 78 | lognormal |

Table 1: Descriptive statistics of the top 13 and cPython (minus setuptools due to small size of contributors) libraries of the Python ecosystem.

The descriptive statistics of the dependency and contributor networks demonstrated that the Python ecosystem is a multi-leveled complex system. The primary diagnostic was the evidence of heavy-tailed distributions in the form of power law and truncated power laws of the dependency network. The application of the statistical approach to each of the top 13 libraries and cPython (as setuptools was removed due to only two contributors) then provided further confirmatory evidence of heavy tailed distributions in both the truncated power law and lognormal form. From this result, the researchers assess that the Python ecosystem is an emergent collective intelligence that evolved over time and found a local optimum to balance the explore-exploit paradigm characteristic of complex systems [1].

# Network Structure

The fact that Python has found a local optimum for its continued development is evident in its success as a programming language. From this success and the fact that the ecosystem has evolved for three decades as a problem solving decentralized social network, this research infers its network structure can provide insights into optimal network structures that must balance both exploration and exploitation, which incorporate modern day knowledge sharing artifacts (e.g. code, git, the Internet). Exploring its network structure then provides insights into network topologies that produce an optimal collective intelligence topology.  To understand this topology, this research explored the network structure in two parts. First, a few descriptive metrics of the dependency graph as it can provide a baseline of the network. Second, an examination of the bipartite and projection of the dependency-contributor graphs by *walking up the tail* of the contributor network.[2]

The dependency network analysis focused on five measures: its connectedness, density, assortativity and its global and local efficiency of information exchange [24]. The dependency graph is a disconnected graph with 475 components, with a giant component that consists of 98% of all nodes or 58,097/59,311. The network is not dense with a density metric of 0.0013.  Its degree assortativity coefficient is -0.1517. This result is consistent with the heavy tailed characteristics of the network where nodes with few connections connect with nodes with many connections.  The dependency network's local efficiency rating is 0.1376 and its global efficiency rating is 0.2676, performing somewhat less efficiently than the WWW or Internet [24] . These metrics provide us with a reasonable understanding of the Python dependency network.

These results, when placed in the context of the ecosystem as a problem solving adaptive network, have interesting implications. The large giant component conforms with the idea that the Python ecosystem is a mostly connected system allowing information to flow nearly anywhere in the system. The assortativity coefficient, which is consistent with the heavy tailed nature of the graph, implies there are only a few solutions everyone needs. This result in combination with the low density of the network is intriguing because it indicates large numbers of possibilities for the network to recombine (or develop new solutions). This has two implications. First, as shown with assortativity, there are only a few widely useful solutions, but there is significant recombination potential for new solutions.  This is consistent with the dynamic of the explore - exploit paradigm and is ubiquitous in biological systems. There are an overwhelming number of ways to explore but only a few will pay off.  The low information exchange scores were surprising because they imply the Python ecosystem in general is not very efficient at exchanging  information, which contradicts the premise of looking at this system. This, however, is discussed more when we look at the contributor network.

Based on the results of the Pratian binning strategy the research then examined the 14 most imported Python libraries and their contributor networks. We did this in two ways: first with a bipartite graph (contributors to libraries) and second with a projection, in that if two contributors contributed to the same library then they were connected. For each projection we also then measured the connectedness, density, assortativity, and global and local efficiency of information exchange. We then examined both the bipartite and projection graph as we *walked up the tail* increasing the size of each graph by adding bins of contributors, determined through the Pratian binning strategy [23].

---

[2] All data and code can be found at
https://github.com/Open-Source-Software-Neighborhood-Watch/dynamics_python_ecosystem

The nature of the graph changes significantly as one expands the bipartite and projection graph to include more of the contributor network. In the extreme tail (or last bin) of the contributor network, all but one of the contributors focus on one library. The result is a highly disconnected graph (14 components of 15 possibilities), a highly assortative graph (with nearly all nodes having one edge) and high local efficiency (0.887) and low global efficiency (0.093). Progressing up the tail to include the first and second bin then results in only five components, lower assortativity, higher local efficiency (0.93) and lower global efficiency (0.08). As one progresses up the distribution to three bins the global efficiency rapidly increases to 0.544 with an extremely high local efficiency at 0.999. Local efficiency then drops slightly to 0.984, while global efficiency continues to rise (table 2). As the dependency network is the reverse projection of the contributor network we can conclude that the local efficiency will continue to drop while the global efficiency will peak and then begin dropping. Although further research is underway to determine where this peak occurs. Placing these results in the context of code development and maintenance then provides insights into problem solving networks.

| Amount of Tail | Nodes | Edges | Components | Density | Assortativity | Local Efficiency | Global Efficiency |
| --- | --- | --- | --- | --- | --- | --- | --- |
| 1 bin | 65 | 191 | 14 | 0.092 | 0.994 | 0.887 | 0.093 |
| 2 bins | 254 | 3568 | 5 | 0.111 | 0.907 | 0.978 | 0.224 |
| 3 bins | 1778 | 268543 | 1 | 0.170 | 0.714 | 0.999 | 0.544 |
| 4 bins | 5702 | 2803769 | 1 | 0.173 | 0.471 | 0.984 | 0.559 |
| 5 bins | 7383 | 4686695 | 1 | 0.172 | 0.547 | 0.986 | 0.572 |

Table 2: Metrics of graph projection of contributor network as one moves up the tail to include more nodes. Bins were developed by using the Pratian binning strategy [23].

The results of the contributor projection graph of the top 14 libraries implies the emergence of specialization within the Python ecosystem that helps form a small world network for efficient information exchange. At the tip of the tail, the contributors are overwhelming specialists (figure 3). As one *walks up the tail*, bridges appear which link Python's critical libraries together, and turn the contributor network into one connected component with near complete local information efficiency. The density also appears to peak and then drop, implying a current maximum of exploited solutions. These multiple peaks are arguably the point where one can see the transition from exploit (maintain processes that work in an efficient manner) to explore (allow individuals to contribute novel ideas). Expanding this research and applying this approach across different OSS ecosystems to assess the dynamics of the explore-exploit trade-off is the focus as the authors work to develop the data necessary for such assessments. For organizations, these explore-exploit network structures can aid in decisions in either finding what the organization feels is its most valuable (e.g. most efficient) processes to identifying how well those processes are linked together by critical nodes.

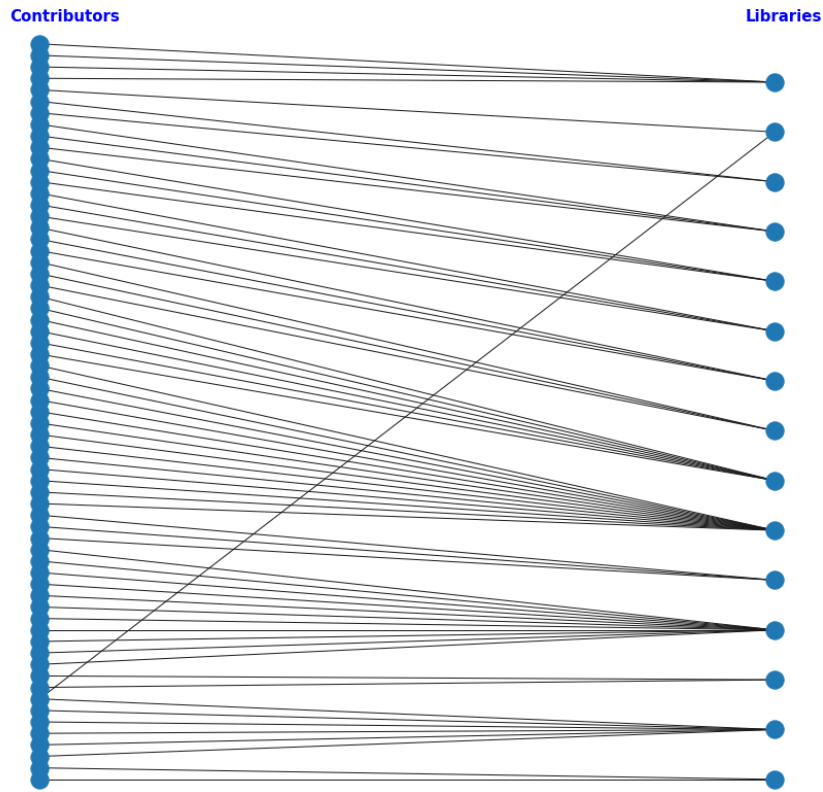

Figure 3: Bipartite graph of the top contributors and their contributions to a library. All but one major contributor focuses on one library. The one exception contributes to both cPython and six, which is a transition library between Python 2 and Python 3.

      The opportunity for further research in this area is substantial. A first step is getting more contributor data to expand this analysis to more of the ecosystem. Second, conducting similar analysis across multiple ecosystems such as Julia, R, and JavaScript to determine if there are any differences or if each ecosystem found a similar local optimum in their network structure. Third, conduct this assessment with longitudinal analysis to see how the network dynamics change over time, ideally including data from the infancy of the ecosystems. A significant challenge is assessing the network as it includes more contributors and libraries. Some thoughts include how can we assess the purpose of smaller libraries that may be seemingly redundant to the more well known libraries? Are new libraries providing potentially better solutions but are not as well known? How do new and novel solutions become popularized? For instance, can we see the rise of machine learning by tracking scikit-learn or Tensorflow and can we identify competitors that were not as successful? Finally, how can we assess the quality of a contribution?  As first shown with the top 14 libraries there is great variance from small and simple with a few contributions to massive with thousands of contributors. Research into any of these areas can provide insights into how organizations and people can function as technology-enabled problem solving networks.

The Python ecosystem represents a well documented problem solving network. Through statistical analysis this research determined it had the critical signatures of a complex system by exhibiting heavy tail distributions in both the dependency graph and each contributor network of the 14 most imported Python libraries. Further analysis of the contributor networks of the 14 most imported libraries revealed the emergence of specialization among contributors, where top contributors focused on one library while major contributors began linking these critical libraries together. The results of the metrics used on the contributor project graph as it was expanded to include more contributors, but *walking up the tail* of the distribution revealed metrics for when the system transitioned from exploiting (high efficiency) to exploring (lower efficiency). The results have implications for organizational dynamics as organizations can examine their networks to identify their most efficient processes and experts for those processes, and who links those processes together. This represents only a small portion of the vast Python ecosystem and much work is still needed. However, it represents possibly significant insights into how humanity and organizations can function more effectively and efficiently as technology enabled problem solving networks.

[1]  S. Kauffman, *At Home in the Universe: The Search for The Laws of Self-Organization and Complexity*. New York: Oxford University Press, 1995.
[2]  J. H. Miller and S. E. Page, *Complex Adaptive Systems: An Introduction to Computational Models of Social Life*. Princton: Princeton University Press, 2007.
[3]  S. E. Page, *The Difference: How the Power of Diversity Creates Better Groups, Firms, Schools, and Societies*. Princeton: Princeton University Press, 2007.
[4]  K. Finley, "Open Source Won. So, Now What?," *Wired*. Accessed: Jul. 17, 2021. [Online]. Available: https://www.wired.com/2016/08/open-source-won-now/
[5]  C. Fershtman and N. Gandal, "Direct and Indirect Knowledge Spillovers: the 'Social Network' of Open-Source Projects," *RAND J. Econ.*, vol. 42, no. 1, pp. 70–91, Mar. 2011, doi: 10.1111/j.1756-2171.2010.00126.x.
[6]  "What is open source?," *Opensource.com*. https://opensource.com/resources/what-open-source (accessed Jun. 25, 2021).
[7]  M. Pittenger, "You are using more open source than you think. It's time to take action," *TechBeacon*. https://techbeacon.com/security/state-open-source-commercial-apps-youre-using-more-you-think (accessed Jun. 25, 2021).
[8]  "Linux Foundation - Decentralized innovation, built with trust," *Linux Foundation*. https://www.linuxfoundation.org/ (accessed Jun. 26, 2021).
[9]  "Foundation Project." https://www.apache.org/foundation/ (accessed Jun. 26, 2021).
[10] H. Simon, *The Sciences of the Artificial*, 3rd ed. Cambridge Massachusetts: The MIT Press, 1996.
[11] "Mining Software Repositories," *Mining Software Repositories*. http://www.msrconf.org/ (accessed Jun. 26, 2021).
[12] A. Decan, T. Mens, and M. Claes, "On the topology of package dependency networks: a comparison of three programming language ecosystems," in *Proceedings of the 10th European Conference on Software Architecture Workshops*, Copenhagen Denmark, Nov. 2016, pp. 1–4. doi: 10.1145/2993412.3003382.
[13] R. Kikas, G. Gousios, M. Dumas, and D. Pfahl, "Structure and Evolution of Package Dependency Networks," in *2017 IEEE/ACM 14th International Conference on Mining Software Repositories (MSR)*, Buenos Aires, Argentina, May 2017, pp. 102–112. doi: 10.1109/MSR.2017.55.
[14] D. Singh Chawla, "The unsung heroes of scientific software," *Nature*, vol. 529, no. 7584, pp. 115–116, Jan. 2016, doi: 10.1038/529115a.
[15] G. Gousios, "The GHTorent dataset and tool suite," in *2013 10th Working Conference on Mining*

*Software Repositories (MSR)*, San Francisco, CA, USA, May 2013, pp. 233–236. doi: 10.1109/MSR.2013.6624034.
[16] J. Creswell, *Research Design: Qualitative, Quantitative, and Mixed Methods Approaches*, 3rd ed. New York: Sage Publications, 2009.
[17] H. Simon, "On a Class of Skew Distribution Functions," *Biometrika*, vol. 42, no. 3/4, pp. 425–440, Dec. 1955.
[18] A.-L. Barabasi and R. Albert, "Emergence of scaling in random networks," *Science*, vol. 286, no. 5439, p. 4, Oct. 1999.
[19] W. Willinger, D. Alderson, J. C. Doyle, and Lun Li, "More "Normal'' Than Normal: Scaling Distributions and Complex Systems," in *Proceedings of the 2004 Winter Simulation Conference, 2004.*, Washington, D.C., 2004, vol. 1, pp. 124–135. doi: 10.1109/WSC.2004.1371310.
[20] C. Cioffi-Revilla, *Introduction to Computational Social Science*. Cham: Springer International Publishing, 2017. doi: 10.1007/978-3-319-50131-4.
[21] A. Clauset, C. R. Shalizi, and M. E. J. Newman, "Power-Law Distributions in Empirical Data," *SIAM Rev.*, vol. 51, no. 4, pp. 661–703, Nov. 2009, doi: 10.1137/070710111.
[22] J. Alstott, E. Bullmore, and D. Plenz, "powerlaw: A Python Package for Analysis of Heavy-Tailed Distributions," *PLoS ONE*, vol. 9, no. 1, p. e85777, Jan. 2014, doi: 10.1371/journal.pone.0085777.
[23] B. Jiang, "Geospatial analysis requires a different way of thinking: the problem of spatial heterogeneity," *GeoJournal*, vol. 80, no. 1, pp. 1–13, Feb. 2015, doi: 10.1007/s10708-014-9537-y.
[24] V. Latora and M. Marchiori, "Efficient Behavior of Small-World Networks," *Phys. Rev. Lett.*, vol. 87, no. 19, p. 198701, Oct. 2001, doi: 10.1103/PhysRevLett.87.198701.